\documentclass[pdflatex,sn-mathphys]{sn-jnl}

\jyear{2021}%

\theoremstyle{thmstyleone}%
\theoremstyle{thmstyletwo}%

\theoremstyle{thmstylethree}%

\begin{document}



\title[Machine Learning Grain Segmentation]{Grain and Grain Boundary Segmentation using Machine Learning with Real and Generated Datasets}



\author[1]{\fnm{Peter} \sur{Warren}}

\author[2]{\fnm{Nandhini} \sur{Raju}}

\author[2]{\fnm{Abhilash} \sur{Prasad}}

\author[2]{\fnm{Shahjahan} \sur{Hossain}}

\author[3]{\fnm{Ramesh} \sur{Subramanian}}

\author[2]{\fnm{Jayanta} \sur{Kapat}}

\author[1]{\fnm{Navin} \sur{Manjooran}}

\author*[2]{\fnm{Ranajay} \sur{Ghosh}}\email{Ranajay.Ghosh@ucf.edu}

\affil*[1]{\orgname{Solve Technology and Research, Inc.}, \orgaddress{\street{3505 Lake Lynda Drive, Suite-200}, \city{Orlando}, \postcode{32817}, \state{FL}, \country{USA}}}

\affil*[2]{\orgdiv{Department of Mechanical and Aerospace Engineering}, \orgname{University of Central Florida}, \orgaddress{\street{4000 Central Florida Blvd}, \city{Orlando}, \postcode{32816}, \state{FL}, \country{USA}}}

\affil*[3]{\orgdiv{Siemens Energy, Inc.}, \orgaddress{\street{11842 Corporate Blvd}, \city{Orlando}, \postcode{32817}, \state{FL}, \country{USA}}}


\abstract{

We report significantly improved accuracy of grain boundary segmentation using Convolutional Neural Networks (CNN) trained on a combination of real and generated data. Manual segmentation is accurate but time-consuming, and existing computational methods are faster but often inaccurate. To combat this dilemma, machine learning models can be used to achieve the accuracy of manual segmentation and have the efficiency of a computational method. An extensive dataset of from 316L stainless steel samples is additively manufactured, prepared, polished, etched, and then microstructure grain images were systematically collected. Grain segmentation via existing computational methods and manual (by-hand) were conducted, to create "real" training data. A Voronoi tessellation pattern combined with random synthetic noise and simulated defects, is developed to create a novel artificial grain image fabrication method. This provided training data supplementation for data-intensive machine learning methods. The accuracy of the grain measurements from microstructure images segmented via computational methods and machine learning methods proposed in this work are calculated and compared to provide much benchmarks in grain segmentation. Over 400 images of the microstructure of stainless steel samples were manually segmented for machine learning training applications. This data and the artificial data is available on Kaggle.}




\maketitle

\section{Introduction}\label{sec1}

Grains are a fundamental hierarchical unit of material organization. For metals, a grain is a region within the material that has a uniform, crystalline structure. Grain size, shape and distribution have a significant effect on the mechanical properties. In general, metals with small, uniform grains are stronger and more ductile than those with large, irregular grains\cite{do2007effect, schempp2013influence, wang1995effect, bai2017effect, voyiadjis2006transient}. In addition to strength and ductility, the grain size and shape can also affect other properties, such as its electrical and thermal conductivity, corrosion resistance, and wear resistance \cite{heo2006influence, uddin2010effect, ali2022computational}. Considering the singular importance of grain structure, its measurement and quantification remains a key objective of materials analysis and characterization. Recent advances in metal additive manufacturing\cite{adam20183d, chen2009modeling, herriott2019multi, raju2022sintering, van2020roadmap}, that depend on layer by layer deposition lead to non-traditional grain structures and hence unique material properties that can be distinct from traditional manufacturing methods\cite{yan2017grain, lin2012microstructure, steinbach2009phase}. Such new technological advances further underscore the need for rapid evaluation of grain structures for predictions and certifications. 

Traditionally grain structures are visible under either optical or Scanning Electron Microscopes (SEM). The grains are then observed via cross sectional imaging and appear like irregular polygons. Although their geometrical structures can be noted manually in a straightforward way, it is too time consuming and impractical for large sample sizes or high throughput testing. Thus, computational techniques that rely on advance image processing algorithms are often employed \cite{bandyopadhyay2020recent, lin2012microstructure, warren2022shrinkage, raju2021material, vayre2012metallic}. However, these techniques suffer from well-known inefficiencies and inaccuracies \cite{yakout2018review, yan2017grain}. These include sensitivity to image quality, parameter tuning requirements, and a lack of standardization \cite{frazier2014metal, tan2020microstructure}. These problems can be addressed using data based Machine Learning (ML) methods. However ML methods are critically dependent on training data sets for their success \cite{perera2021optimized}. Such training data sets are often tedious, expensive and difficult to obtain, severely restricting the use of machine learning. 

\begin{figure}[ht]%
\centering
\includegraphics[width=0.8\textwidth]{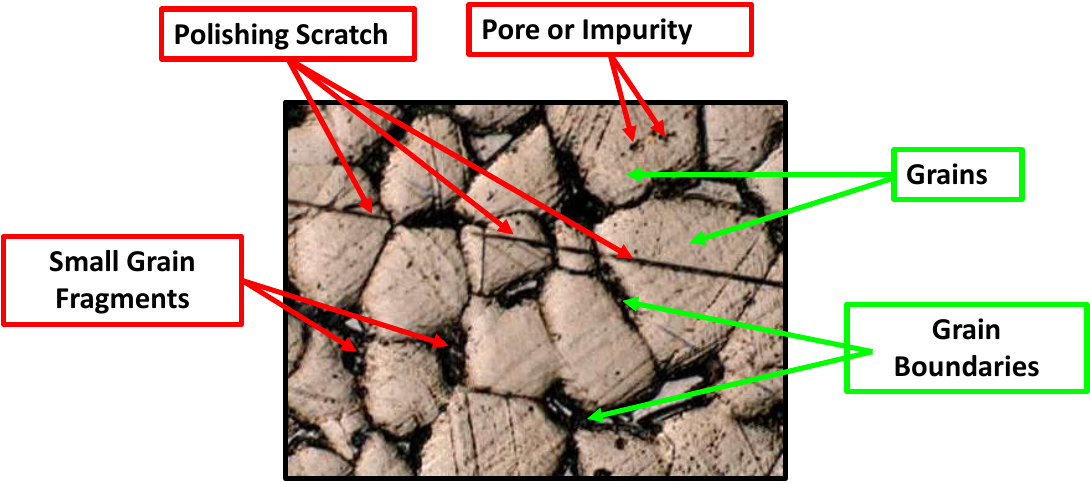}
\caption{An example image of a metallic microstructure which illustrates the challenges during classification of pixels into either grains or grain boundaries. These challenges include: polishing scratches, pores, impurities, and small grain fragments.}
\label{Chal}
\end{figure}

This problem is addressed in this paper by developing a new approach whereby a hybrid experimental and Voronoi cell generated images are used as training set. Artificial grain images are produced by exploiting the geometries that occur in a Voronoi cell plot, and can therefore be controlled and manipulated through the mathematical properties of the Voronoi method. In addition, simulated defects such as pores and polishing scratches were added, and guassian noise is added in a myriad of ways to create a finalized and realistic grain image. The artificial grain generation method in this work produces equiaxed grains, which was observed on the experimentally gathered data. This is not a limitation as this method could easily be tuned to create other grain structure patterns as well. The artificial and authentic (naturally occurring) training data was used to train CNN’s to perform grain segmentation. Once segmentation is performed the accuracy was calculated from the geometric grain measurements and from the pixel segmentation accuracy. The grain and grain boundary segmentation process can be challenging due to pores, impurities, polishing scratches, and small grain fragments in the images. An illustrative example of these challenges
is shown in Fig \ref{Chal}. Artifially generated data methods are often computationally expensive, and this novel method significantly reduces computation by generating the underlying geometric pattern and adding intricacies directly through mathematical pattern manipulation in euclidean space.

Traditional techniques such as the line intercept and planimetric methods have been widely used for this purpose \cite{heyn1925physical, subcommittee1996standard}. In this study, a large dataset of metallic grains from cubical samples of stainless steel 316L additively manufactured using an InnoventX ExOne printer (Binder Jet Technology) and an optical microscope is collected and compiled. After printing, sintering, polishing, and etching the samples, a total of 640 images were collected and manually segmented as the baseline for evaluating various segmentation methods. This dataset provides a valuable resource for testing and comparing different machine learning techniques for grain boundary segmentation \cite{PW1, PW3}.

The rest of the paper is organized as follows, a description of the traditional image segmentation techniques is given in Section 2. These include manual thresholding, gradient based approach, holistically nested edge detection (HED), and manual segmentation. A description of how methodology accuracy is calculated, and accuracy of segmentation is calculated is provided in Section 3. The design and implementation of the novel artificial grain generation method developed in this work is provided in Section 4. The CNN architecture and the composition of the various training sets is provided in Section 5. Finally the results and discussion is given in Section 6, and then the conclusion in Section 7.

\section{Image Segmentation Techniques}\label{sec2}

 The section of the paper will explain the four most common ways to segment an image into grains and grain boundary. The four ways are: (1) Manual Thresholding (2) Gradient Based Methods (3) HED method (4) Manual Segmentation. In practice these methods can be combined in a variety of ways to segment and image. The HED method is an example of a neural network style approach. There are many new developing methods that use neural networks to detect edges of an image, HED is just the most well known and popular at the time of writing this paper. The HED method was developed specifically for detecting the edges of an image in any image.

\subsection{Manual Thresholding}

\begin{figure}[ht]%
\centering
\includegraphics[width=1\textwidth]{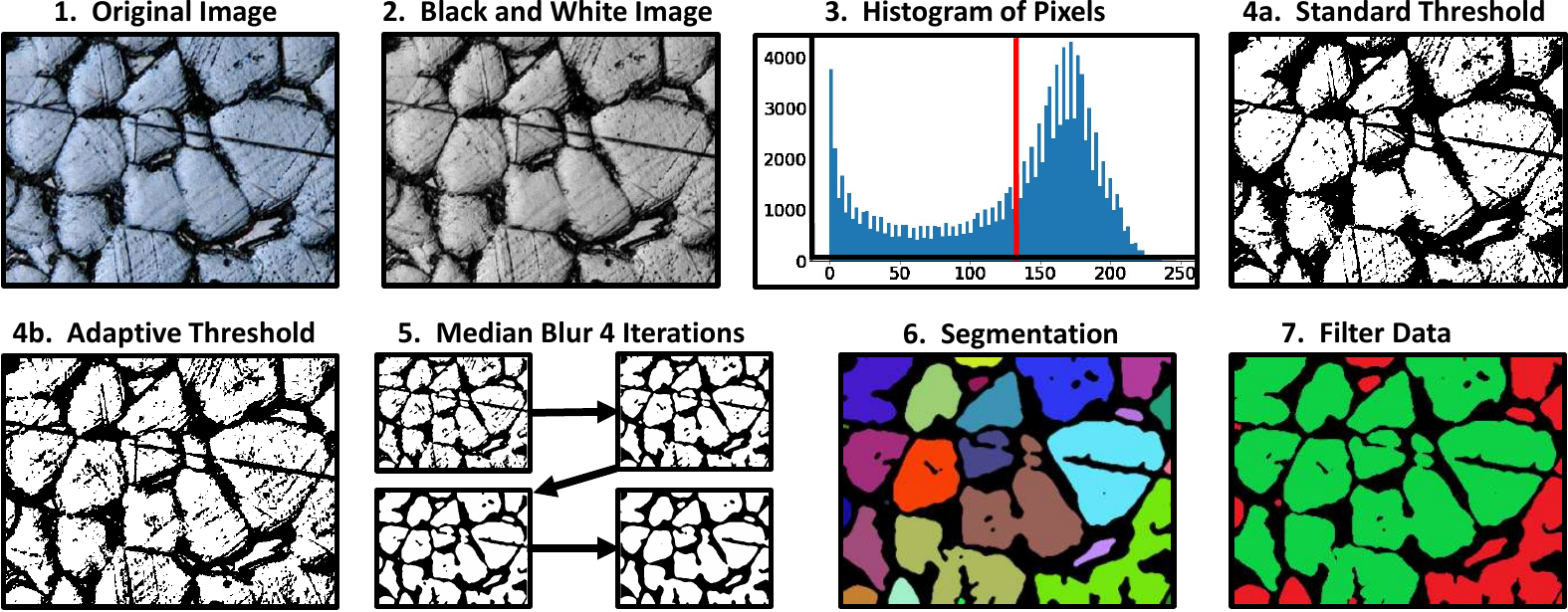}
\caption{A step by step example of the manual thresholding technique being applied to a metallic microstructure image with equiaxed grains boundaries.}
\label{Thresh}
\end{figure}

The coding for the manual thresholding was done using OpenCV \cite{opencv_library}. All of the code for this work is available on github and kaggle \cite{PW}. A step by step approach to manual thresholding can be seen in Fig \ref{Thresh}. Step 1 is to collect the microstructure image. The material in this work is 316L stainless steel printed on an EXone printer from an optical microscope at 500X. Step 2: Convertion of image to black and white. Step 3 is a pixel intensity evaluation for pixel cutoff determination. The histogram shown for step 3 is for this purpose, and the X-axis is pixel intensity and the Y-axis shows the number of times that pixel intensity has occurred. The red line on this histogram shows the selected thresholded value, also called the pixel cutoff value. Step 4a and step 4b are both examples of thresholding. Step 4a is an example of standard thresholding where all of the pixels ($>$130) are set to 255 (white) and all pixels ($\leq$130) are set to 0 (black). The value 130 is shown in red on the histogram. Step 4b is an example of an adaptive gaussian threshold. The threshold value is a gaussian-weighted sum of the neighbourhood values minus the constant C. The neighborhood values are a block of pixels, for this case it was (55x55), and the constant subtracted was 2. 

In step 5 a median blur is applied 2 times to a 5x5 pixel area, and then 2 more times to a 7x7 pixel area. In step 6 image segmenation is applied, and the grains are seperated from the grain boundaries. In step 7 the grains are filtered, usable grains are shown in green and unusable grains are shown in red. Unusable grains are grains that are too small or grains that have too much perimeter per area. Too much perimeter per area indicates the grain did no get properly segmented during the total process. Essentially two or more grains are connected by a small "neck" region. 

Image segmentation via threshold is a simple but highly customizable process. There are several steps in this process where the user has to select a value and check the results. For example, threshold cutoff, block size for blurring, block size for the adaptive gaussian method, etc. All of these values lead to different results, so the user must pick the one that looks the best in order to automate the grain segmentation process. Different materials, imaging equipment, or image settings can effect the images, and could easily require different variables for optimal segmentation.

\subsection{Gradient Based Methods}

 The gradient based approach identifies sharp changes in pixel intensity, and can classify those areas as an edge. The first and second derivatives of pixel intensity as it relates to pixel position are used to determine the gradients. The method used in this work is the "Canny Edge Detection Method," developed by John Canny \cite{canny1986computational}. First the image is converted to greyscale and then gaussian filters are used on the pixel values to slightly blur the image. This will reduce the amount of predicted edges in the result, which is helpful because the gradient based approach tends predict too many edges. After the gaussian filters, the pixel intensity gradient is calculated (first and second derivatives). Lower changes in pixel intensity are filtered out, unless they directly connected to a high change in pixel intensity. 

\begin{figure}[ht]%
\centering
\includegraphics[width=0.8\textwidth]{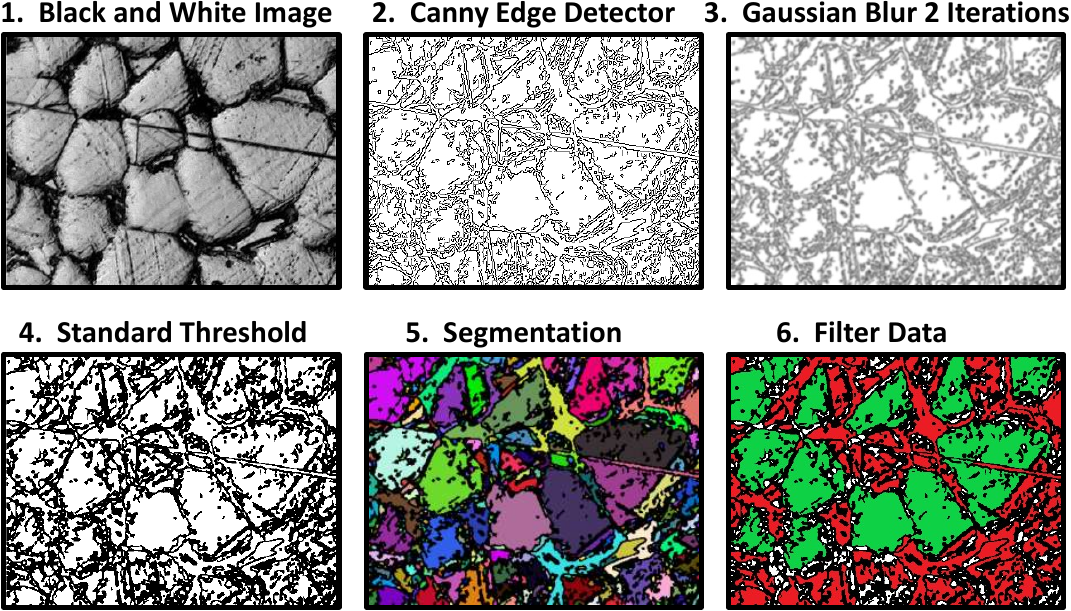}
\caption{A step by step example of the gradient based technique being applied to a metallic microstructure image with equiaxed grains boundaries.}
\label{Grad}
\end{figure}

The application of the Canny Edge Detector can be seen in Fig \ref{Grad}. First the image was converted to black and white. Next the Canny Edge Detector was used. In the third step, two Gaussian blur iterations were applied to a 3x3 pixel block. In the fourth step, the image is thresholded into 0 and 255 (black and white). The threshold value was very high (200), meaning most gray pixels were set to black. Step 5: segmentation occurs. Step 6: Filtering of the data occurs. The same filters were used as the previous section. Too much perimeter per area were filter out, and too little a grain size were also filter out. Image segmentation via gradient detection is a highly customizable process and optimal parameters can be difficult to determine. Small adjustments to pre and post filters will cause significant changes in the final predictions. The engineer or user of this method must tailor the settings and parameters of the algorthm for specific materials or image settings.

\subsection{Holistically Nested Edge Detection}

The next detection method is deep learning-based method for edge detection in images. The architecture of the HED utilizes a fully convolutional network, which was trained end-to-end on a large dataset of images with ground truth edge maps.\cite{xie2015holistically} CNN's are particularly well-suited for tasks involving image processing and analysis, such as object detection, image classification, and image segmentation \cite{o2015introduction, albawi2017understanding, warren2021rapid}. 

\begin{figure}[ht]%
\centering
\includegraphics[width=0.8\textwidth]{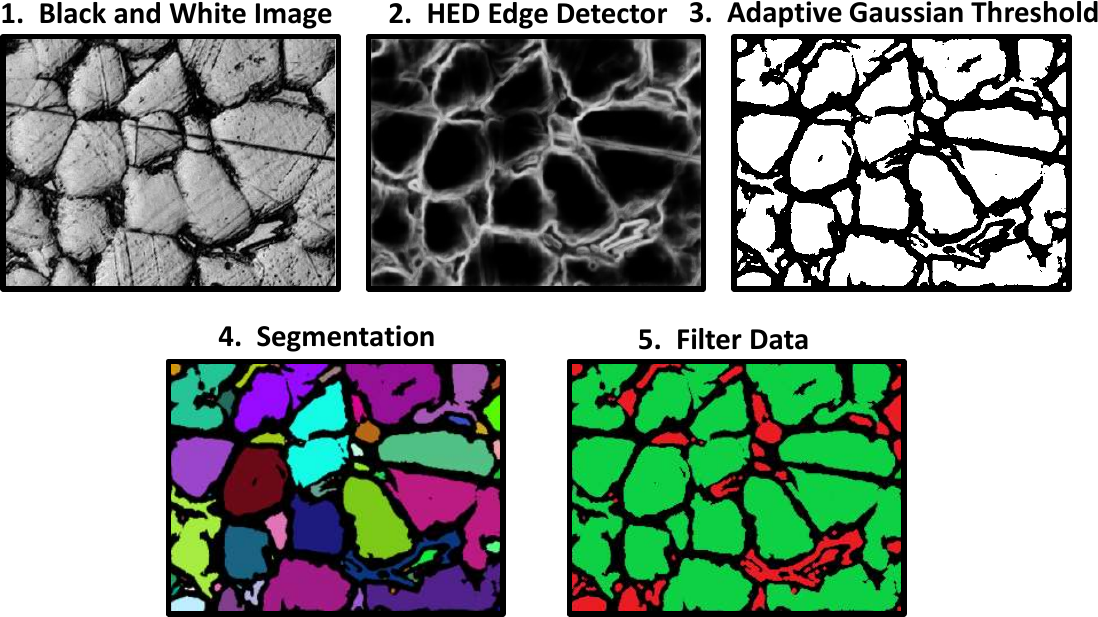}
\caption{A step by step example of the (HED) Holistically Nested Edge Detection neural network technique being applied to a metallic microstructure image with equiaxed grains boundaries.}
\label{HED}
\end{figure}

In Fig \ref{HED}, a clear step-by-step guide for utilizing the HED approach is provided.  First the image is converted to black and white. Next it is fed into the HED model to determine the edges. The third step is to apply an adaptive guassian threshold to convert all pixels to either 0 or 255 (white or black). The final two steps are segmentation and data filtration. The HED model outputs a pixel intensity that represents the confidence that the pixel is an edge or not an edge. That is why step 3 (adaptive thresholding) is required. The HED is highly effective at detecting grain boundaries, but its suitability as an end-to-end solution is limited due to its tendency to also detect undesired features such as pores and polishing scratches. This is primarily because HED lacks the ability to distinguish between different types of edges, relying on low-level image features that generalize its edge detection capabilities to various image characteristics.

\subsection{Manual Segmentation}

\begin{figure}[ht]%
\centering
\includegraphics[width=1\textwidth]{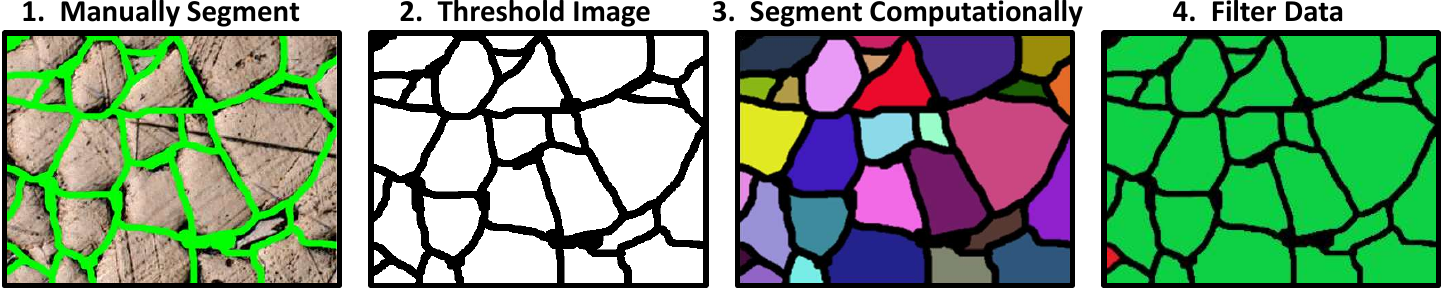}
\caption{A step by step example of the manual segmentation being applied to a metallic microstructure image with equiaxed grains boundaries. This was done by hand, by a human.}
\label{Manual}
\end{figure}

The most arduous way to segment the grains of image is the manual segmentation. This is done by the user manually annotating the grains or grain boundaries. For this work it was completed by marking the grain boundaries in a specific color, and then thresholding the image so that all annotated pixels are 0 (black) and all others are 255 (white). After that the same process of segmenting the image computationally and filtering out small particles occurs. This process is shown in Fig \ref{Manual}. All inaccuracies that occur in this method are from human error. Polishing scratches and pores must be avoided during annotation of the grain boundaries. This method cannot be automated, each image requires human attention. For this work all 480 images of grains were painstakingly manually segmented, and this was completed to allow the selected machine learning methods to train and replicate the process. All of the previously mentioned segmentation methods/algorithms were used (HED, Gradient, Threshold) on the dataset to provide benchmarks for each of the methods and the proposed methods in this work. All of this data is available on Github and Kaggle.

\section{Grain Measurement and Accuracy Quantification}\label{sec3}

There are multiple ways to quantify how accurate the grain boundary segmentation methods perform. Accuracy quantification can be approached from the perspective of a material scientist or from the perspective of a computer vision scientist. Both approaches will be conducted in this work. The material scientitst is interested in the resulting dimensional grain data. Traditional grain measurements techniques are used once the image is segmented. The line intercept (Heyn method) and the planimetric method (Saltykov method) are both used to measure the grain dimensions. These two industry standard methods will be explained in the following sections. The computer vision scientist is interested in the pixel segmentation accuracy. The current standard is the dice score or the Intersection of Union (IoU) score. These two methods are frequently used in computer vision to determine how accurate a nueral network is at segmentation, and they will be explained in this section as well. 

\subsection{Line Intercept Method}

The grain boundary image dataset used in this study consists of 480 images, each with dimensions of 400 x 300 pixels. The grain intercept method, initially developed by Heyn and widely employed for rapid grain size measurements \cite{heyn1925physical}, involves drawing a line across the image and tallying the number of grains intersected by the line. The average grain size is then determined by dividing the total number of grains by the line length, as shown in Equation \ref{line}. To automate this process and conduct measurements at each pixel, code was implemented for all the segmentation techniques outlined in Section 2, resulting in a total of 144,000 X-direction measurements and 192,000 Y-direction measurements. Table \ref{Line_Table} presents the average grain size and standard deviation for both the horizontal and vertical directions. Although the grain intercept method provides a substantial number of measurements, it is inherently limited as a 1-dimensional approximation of a complex 3-dimensional structure. Nonetheless, its continued utilization in current research reinforces its relevance in this study.

\begin{equation}
Grain\ Size = \frac{Number\ of\ Grains\ in\ Contact\ With\ Line}{Total\ Length\ of\ the\ Line}
\label{line}
\end{equation}

\begin{table}[ht]
\begin{center}
\begin{minipage}{\textwidth}
\caption{Line Intercept Grain Measurements}\label{Line_Table}%
\begin{tabular}{lccccc}
\toprule
Direction & Manual & Gradient & HED & Manual \\
 & Threshold  &  Method & Method & Segmentation \\
\midrule
X-Direction   & $41.45 \pm 6.42$ & $20.86 \pm 10.90$  & $33.05 \pm 6.53$  & $36.87 \pm 4.84$ \\
\midrule
Y-Direction   & $42.28 \pm 7.59$ & $22.64 \pm 5.99$  & $34.02 \pm 6.80$  & $37.40\pm 5.44$ \\
\botrule
\end{tabular}
\end{minipage}
\end{center}
\end{table}

\subsection{Planimetric Method}

The planimetric method, a 2-dimensional approach, enables the extraction of size and shape data regarding the grains. Extensive data from the segmented dataset was collected, which has been categorized into size data (Table \ref{Size_Table}) and shape data (Table \ref{Shape_Table}). These datasets encompass all the segmentation methods listed in Section 2. The planimetric size data comprises four measurements. The first on is the total grains metric which represents the overall count of grains across all images. The second measurement is the average individual grain area, accompanied by its standard deviation. The third and fourth entries in Table \ref{Size_Table} correspond to the total grain area per image and the total grain boundary area per image. These two values can be computed interchangeably using Equation \ref{GB2GArea}.

\begin{equation}
Total\ Image\ Area = Total\ Grain\ Area + Total\ Grain\ Boundary\ Area
\label{GB2GArea}
\end{equation}

\begin{equation}
Circularity = \frac{4\pi A}{P^2} 
\label{Circ}
\end{equation}

\begin{table}[ht]
\begin{center}
\begin{minipage}{\textwidth}
\caption{Planimetric Size Measurements}\label{Size_Table}%
\begin{tabular}{cccccc}
\toprule
Measurement & Manual & Gradient & HED & Manual \\
& Threshold  &  Method & Method & Segmentation \\
\midrule
Total Grains  & 9,575 &  24,740  & 15,297  & 11,002 \\
\midrule
Individual Grain & $844 \pm 1,314$ & $ 248 \pm 1033$  & $488 \pm 823$  & $1068 \pm 929$ \\
Area Average &   $\mu m^{2}$ & $\mu m^{2}$ & $\mu m^{2}$ &  $\mu m^{2}$  \\
\midrule
Total Grain & $53.3\%$ & $40.5\%$   & $49.3\%$  & $77.5\%$  \\
Area $\%$ &        &    &    &   \\
\midrule
Total Grain & $46.7\%$ & $59.5\%$   & $50.7\%$  & $22.5\%$ \\
Boundary Area $\%$ &        &    &    &   \\ 
\botrule
\end{tabular}
\end{minipage}
\end{center}
\end{table}

The planimetric shape data encompasses five distinct measurements. Circularity represents a geometric property of a shape, ranging between 1 and 0, where 1 denotes a perfect circle and 0 signifies the theoritical opposite of a perfect circle. The closer the value is to 0, the rougher the 2-dimensional shape becomes. The equation for circularity can be seen in Equation \ref{Circ}. Another measurement is the average maximum diameter, which represents the maximum distance from one side of the grain to the other, passing through the centroid of the grain. The subsequent three measurements pertain to the X and Y directions. The first is the average width of the grains, denoted as the average X-diameter. The next measurement is the average diameter in the Y-direction, also known as the average height. The final value is the aspect ratio, obtained by dividing the average X-diameter by the average Y-diameter, providing a measure of the width-to-height relationship.

\begin{table}[ht]
\begin{center}
\begin{minipage}{\textwidth}
\caption{Planimetric Shape Measurements}\label{Shape_Table}%
\begin{tabular}{cccccc}
\toprule
Measurement & Manual & Gradient & HED & Manual \\
& Threshold  &  Method & Method & Segmentation \\
\midrule
Circularity & $ 0.55 \pm 0.18 $ & $0.28 \pm 0.15$  & $ 0.50 \pm 0.18$  & $0.61 \pm 0.12$ \\
\midrule
 Average Max & $39.31 \pm 30.69$ & $25.14 \pm 18.60$  & $30.57 \pm 22.84$  & $47.38 \pm 20.52$ \\
Diameter &        &    &    &   \\
\midrule
Average & $32.18\pm 25.39$ & $22.64 \pm 15.90$  & $24.81\pm 19.51$  & $39.07 \pm 18.49$ \\
X-Diameter &        &    &    &   \\
\midrule
Average & $32.94 \pm 26.70$ & $22.64 \pm 15.40$  & $25.47\pm 19.76$  & $39.06 \pm 18.97$ \\
Y-Diameter &        &    &    &   \\
\midrule
Aspect Ratio & $1.09 \pm 0.55$ & $1.06 \pm 0.53$  & $1.07 \pm 0.51$  & $1.11 \pm 0.58$ \\
X/Y &        &    &    &   \\
\botrule
\end{tabular}
\end{minipage}
\end{center}
\end{table}

\subsection{Evaluation of Segmentation}

For assessing the segmentation model accuracy, the dice score and the IoU (Intersection over Union) score will be employed. When classifying pixels in a segmented image, there are four possible outcomes: true positives (TP), false positives (FP), true negatives (TN), and false negatives (FN). The dice equation, represented by these variables, can be seen in Equation \ref{dice}. Additionally, the probabilistic version of the equation is provided in Equation \ref{diceAB}. These equations serve as quantitative measures to evaluate the performance of the segmentation models.

 \begin{equation}
Dice = \frac{2TP}{2TP + FN + FP}
\label{dice}
\end{equation}

\begin{equation}
Dice(A,B) = \frac{2*{\lvert} A \cap B {\rvert}}{\lvert A \rvert + \lvert B \rvert}
\label{diceAB}
\end{equation}

To facilitate implementation, it is crucial to establish and indicate a ground truth. In this study, the manually segmented images serve as the ground truth, with the only potential error arising from human inaccuracies during the grain boundary sketching process. Table \ref{diceTable} showcases the comparison of Dice and IoU scores for the manually thresholded method, the gradient method, and the HED method. These scores can serve as benchmarks, alongside other grain properties, to quantify the accuracy of the machine learning methods developed in this research.

\begin{table}[ht]
\begin{center}
\begin{minipage}{\textwidth}
\caption{Planimetric Shape Measurements}\label{diceTable}%
\begin{tabular}{cccccc}
\toprule
Scoring Method & Manual Threshold & Gradient & HED \\
\midrule
Dice & $ 0.55 \pm 0.18 $ & $0.28 \pm 0.15$  & $ 0.50 \pm 0.18$\\
\midrule
IoU & $39.31 \pm 30.69$ & $25.14 \pm 18.60$  & $30.57 \pm 22.84$\\ 
\botrule
\end{tabular}
\end{minipage}
\end{center}
\end{table}

\section{Artificially Generated Grains}\label{sec4}

In this work, a novel approach was developed to generate artificial grains and grain boundaries using a Voronoi tessellation pattern. By leveraging the Voronoi tessellation pattern, specifically tailored to meet the desired conditions, a pattern closely resembling equiaxed grain boundaries was achieved. To ensure realism and account for practical challenges, artificial noise was intentionally introduced to simulate pores, scratches, and various image quality issues. This innovative Voronoi grain method, developed within the scope of this research, enables the generation of representative synthetic grain structures for further analysis and evaluation.

\subsection{Voronoi Tessellation}

\begin{figure}[ht]%
\centering
\includegraphics[width=1\textwidth]{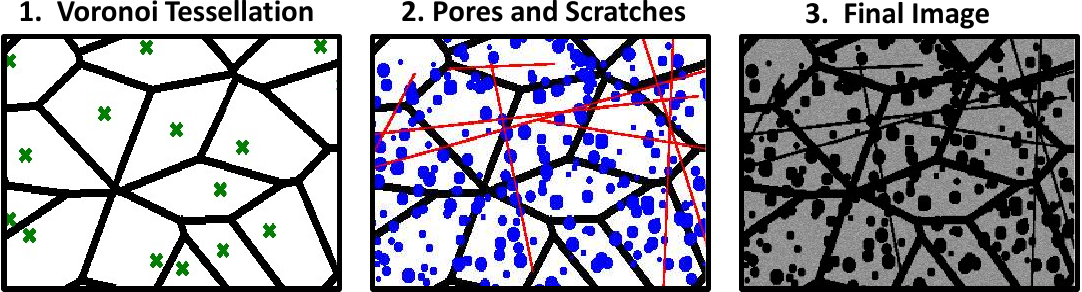}
\caption{Artificial grain generation via Voronoi Tessalation pattern: (1) Randomly generated voronoi tessalation pattern with centroids marked in green. (2) Polishing scratches marked in red and large and small pores marked in blue. (3) Gaussian Noise applied to simulate grain color.}
\label{Voro}
\end{figure}

A Voronoi tessellation pattern is generated by dividing a plane into regions based on the distances to specific points in the plane \cite{voronoi1908nouvelles}. Each point, also referred to as a seed, corresponds to a region comprising all the areas in the plane that are closer to that point than to any other. This partitioning of the plane into regions defined by proximity to the seeds forms a Voronoi diagram. An example of a Voronoi diagram, utilized in this work, is depicted in Figure \ref{Voro}. The size of the Voronoi cells is determined by the area measurements obtained in the previous section. The average area of the Voronoi cells is approximately equal to the reciprocal of the point density. For instance, if the point density is 10 points per unit area, the average cell area would be approximately 0.1 units. Based on this calculation, the average grain (cell) area for the generated data was set to 1000 $\mu m^{2}$. Comparatively, the average individual grain area for the manually segmented images was found to be 1068 $\mu m^{2}$, so 1000 $\mu m^{2}$ was selected to closely represent the actual data.

\subsection{Artificial Noise}

To accurately replicate the challenges observed in the images, noise was intentionally introduced into the Voronoi tessellation pattern. The main challenges encountered in the images include polishing scratches and pores, as depicted in Figure \ref{Chal}. To incorporate scratches, random lines were plotted within the Voronoi graph, while pores were simulated by randomly placing dots of varying sizes. An illustrative example showcasing these defects can be seen in Figure \ref{Voro}. While these defects could be mitigated through more meticulous polishing techniques and improved etching methods, achieving perfection in these complex processes can be challenging, especially for certain metals. Therefore, employing machine learning methods to automatically disregard these defects represents a viable approach to mitigating their undesired effects during the measurement process.

The overall plot area ranges from 0 to 800, but the actual image spans from 200 to 600. Consequently, a 400x400 pixel image, identical to the experimental data, is created. This arrangement allows for the possibility of scratches either traversing the entire sample image or being confined within the image boundaries. During the generation of the artificial data, the quantity of pores and scratches was deliberately varied, and this will be further explained in the subsequent section.

\subsection{Artificial Dataset}

The quantity of scratches within the entire plot area was determined using a uniform random number generator ranging from 2 to 10. The starting and stopping locations of the scratches were also randomly assigned using uniform random distributions for both the X and Y coordinates. Moreover, the number of randomly sized pores was generated using a uniform random number between 2000 and 4000. Additionally, random black pixels were inserted into the images, with the quantity determined by a uniform random number between 10,000 and 20,000. To impart a gray color to the grains, a subtraction operation was applied to all white pixels (255) using a normally distributed Gaussian distribution centered around $100 \pm 10$. Detailed numerical values for all of noise distributions can be found in Table \ref{AllTab}. The objective is to train a CNN to accurately predict the mask, representing the initial noiseless Voronoi tessellation, given a noisy Voronoi tessellation. Although a CNN can readily accomplish this task, additional noise will be introduced to enhance the robustness of the model.

\begin{table}[ht]
\begin{center}
\begin{minipage}{\textwidth}
\caption{Applied Noise Distribution Table}\label{AllTab}%
\begin{tabular}{cccccc}
\toprule
Noise Type & Distribution/Application \\
\botrule
Added Black Pixels &  $\mathcal{U}\sim$ (10,000, 20,000) \\
\midrule
Added Black Dots & $\mathcal{U}\sim$ (1,000, 2,000) \\ 
\midrule
Dot Size (Radius) & $\mathcal{U}\sim$ (5, 25) \\
\midrule
Gaussian Noise Subtracted & $\textbf{N} \sim$ (100, 10) \\ 
\midrule
Total Scratches & $\mathcal{U}\sim$ (2, 10) \\
\midrule
Scratch Start and End Point & $\mathcal{U}\sim$ (0, 800) \\ 
\midrule
Total Area Plotted & X and Y : (0, 800) \\
\midrule
Area in Image (From Original Plot) & X and Y : (200, 600) \\ 
\botrule
\end{tabular}
\end{minipage}
\end{center}
\end{table}

\begin{figure}[ht]%
\centering
\includegraphics[width=1\textwidth]{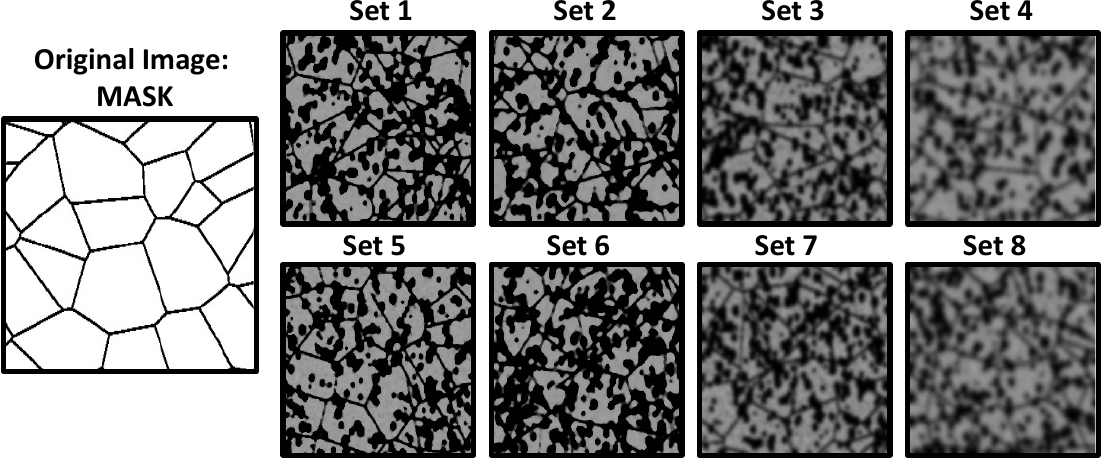}
\caption{Artificial grain generation method: The original mask is on the left hand side of the figure and all eight of the different sets of generated grains are labeled Set 1-8. The method of noise generation and application for each set is given in Table \ref{DataGenTab}}
\label{Generated}
\end{figure}

To enhance the training process and develop a more robust CNN for image segmentation, the next round of noise was introduced by exploiting the limitations of two segmentation techniques mentioned in Section 2: Manual Thresholding and Gradient-Based Method. This approach aims to leverage the deficiencies of these techniques to further train the model. Specifically, the median blur technique described in Section 2.1 and the Gaussian blur technique discussed in Section 2.2 were implemented. Additionally, the inverse of an adaptive threshold, which is similar to an erosion technique, was applied. These techniques were utilized in various iterations and combinations to create a diverse range of noise patterns. The specific application method can be found in Table \ref{DataGenTab}. For a visual representation of the generated noise applied to an original Voronoi tessellation for each set in Table \ref{DataGenTab}, refer to Figure \ref{Generated}.

\begin{table}[ht]
\begin{center}
\begin{minipage}{\textwidth}
\caption{Image Distortion Applications Table}\label{DataGenTab}%
\begin{tabular}{cccccc}
\toprule
Set of 100 & Erosion and Dilation         & Gaussian Blur    &   Median Blur  \\
Images 256x256  & Kernel [3,3]    & Kernel [15,15]   &  Kernel [5,5] \\
\botrule
    1       &      YES      &       NO      &  YES (1 iteration) \\
\midrule
    2      &      YES      &       NO      &  YES (2 iteration) \\
\midrule   
    3       &      YES      &       YES (1 iteration)      &  NO \\
\midrule   
    4       &      YES      &       YES (2 iteration)      &  NO \\
\midrule   
    5       &      NO        &       NO      &  YES (1 iteration) \\
\midrule   
    6       &     NO         &       NO      &  YES (2 iteration) \\
\midrule   
    7       &      NO         &       YES (1 iteration)      &  NO          \\
\midrule   
    8       &     NO         &       YES (2 iteration)      &  NO           \\
\botrule
\end{tabular}
\end{minipage}
\end{center}
\end{table}

\section{Machine Learning Methods}\label{sec5}

The chosen machine learning approach for grain and grain boundary identification is a Convolutional Neural Network (CNN) with a U-Net architecture. The details of the U-Net architecture will be discussed in the  section 5.1: U-NET. Multiple training methods were employed and will be elaborated upon in the section covering 5.2: Training Styles.

\subsection{U-Net}

The U-Net architecture, introduced in 2015 by Ronneberger et al. \cite{ronneberger2015u}, has emerged as a powerful framework for pixel segmentation tasks. Its design is particularly well-suited for this purpose as it effectively captures both local and global features present in the input image. One key advantage of U-Net is the incorporation of skip connections, enabling the model to leverage both high-level and low-level features during the segmentation process. In many image classification problems, both levels of features contain crucial information necessary for accurate classification.

In the current era of Artificial Intelligence growth, numerous variations of the classic U-Net model have been proposed. However, to maintain consistency and facilitate benchmarking, we have opted to employ the standard U-Net architecture in this study. Comparing the performance of different models can be easily accomplished since all the relevant data is publicly available.

\begin{figure}[ht]%
\centering
\includegraphics[width=1\textwidth]{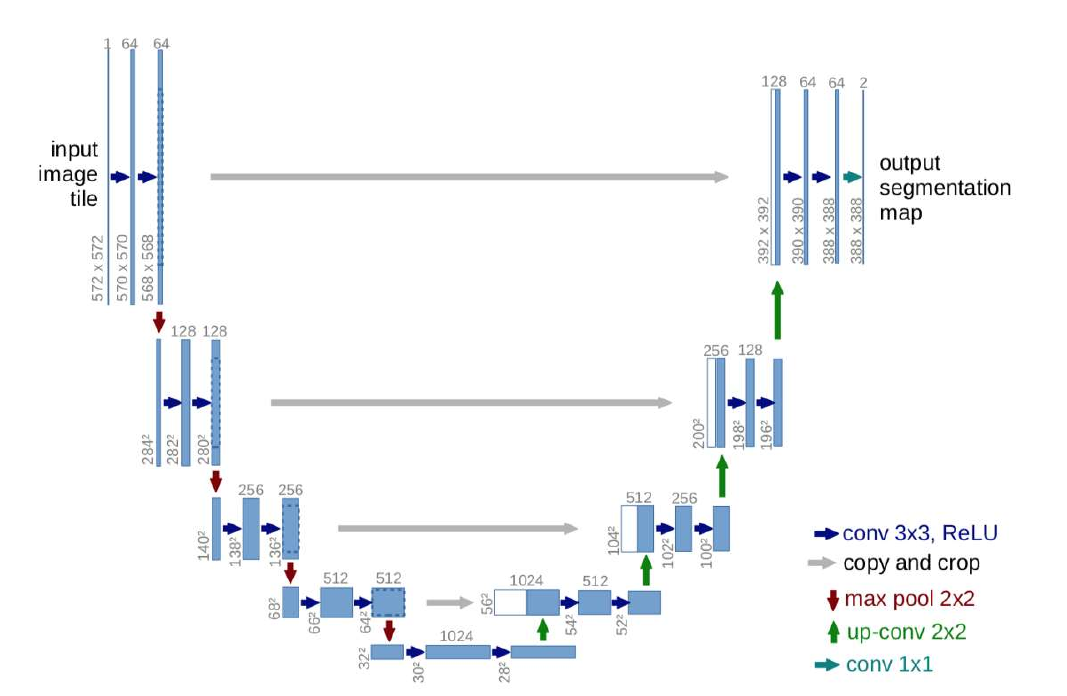}
\caption{This is a diagram of a UNET from the First and Original paper on the UNET. caption\cite{ronneberger2015u}}
\label{UNET}
\end{figure}

For the activation functions within the layers, the Rectified Linear Unit (ReLU) function was utilized. The final activation function for the classification task was the sigmoid function. The optimizer function employed was ADAM, while the loss function used was binary cross-entropy. During training, a batch size of 16 was utilized, and the training process consisted of 50 epochs. A validation split of 10$\%$ was employed to monitor the model's performance during training. The layout of the U-Net architecture employed in this study is depicted in Figure \ref{UNET}.

\subsection{Training Styles}

The dataset was split into 80$\%$ for training and 20$\%$ for testing. Two crucial aspects of the training process are determining the training data and defining the desired outcome, referred to as the mask. The mask represents the specific pixels classified, in this case classified as gran boundaries. In this study, the mask (grain boundary classified pixels) are assigned a value of 0 (black), while all grains are assigned a value of 255 (white). The masks used in this work were primarily manually segmented by human experts, except for the Voronoi-generated data, where the original unaltered Voronoi diagram serves as the mask. An example of such a mask is illustrated in Figure \ref{Voro}.

\begin{figure}[ht]%
\centering
\includegraphics[width=1\textwidth]{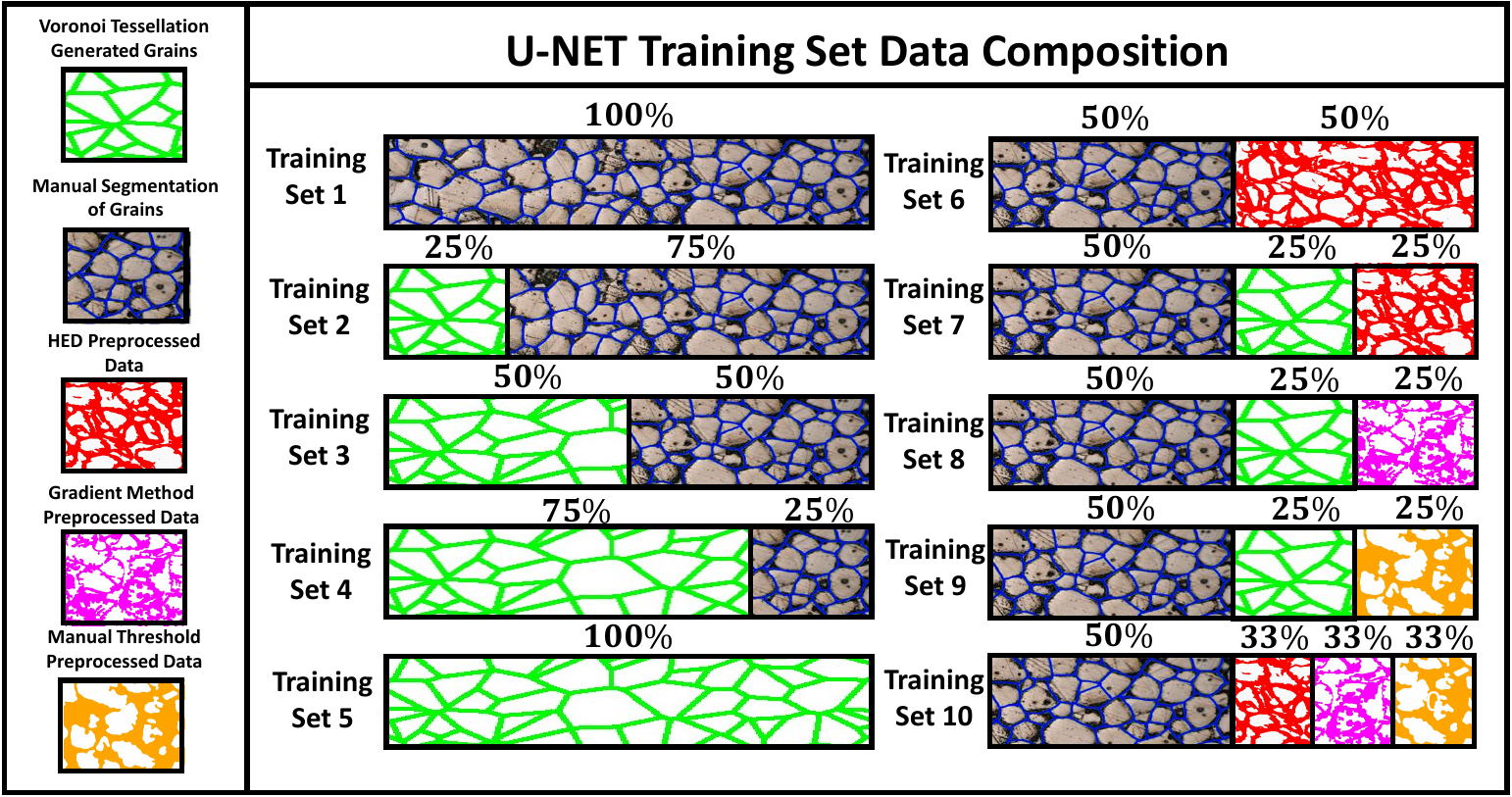}
\caption{A visual grouping of the various sets of data used for training neural networks in this work. The training sets are composed of either manually segmented data, artificially generated data, or preprocessed data through the HED method, Gradient Method, or Manual Threshold method}
\label{train}
\end{figure}

The total composition of the training data is a critical aspect to consider. The obvious approach is to train  the model solely on manually segmented data. However, the question arises: would incorporating the Voronoi tessellation-generated data enhance the model's robustness? This question will be addressed in the results section. The different compositions of the training sets are depicted in Figure \ref{train}. The first five compositions involve varying percentages of generated and manually segmented data, allowing for a comparison between the Voronoi method, the Manual method, and a combination of both.

Training sets 6-10 follow a more intricate approach. Traditional segmentation methods from Section 2.1, 2.2, and 2.3 were utilized to preprocess the data. This means that the mask remains the same (manually segmented image), but the original grain image undergoes modifications through the application of the manual threshold method, the HED method, or the gradient method. Subsequently, the U-Net model is trained to map from the manually segmented image to the mask (manually segmented image). This approach was employed to explore whether it would lead to a more robust model and what effect the incorporation of this data would have on the accuracy of the model.

\section{Results and Discussions}\label{sec6}

In Section 3: Grain Measurement and Accuracy Quantification, two methodologies were discussed to assess the accuracy of the results. The first approach utilizes a commonly used metric for segmentation accuracy called the Dice score. This quantification method provides a measure of how well the predicted segmentation masks align with the ground truth masks. The second approach focuses on evaluating the accuracy of dimensional grain data measurements. This involves comparing the measured grain properties obtained from the predicted masks with the ground truth measurements. Both of these evaluation methods were implemented and will be discussed in detail in the subsequent sections. In addition to the quantitative results, the qualitative findings of this work will also be presented and discussed.

\subsection{Dice Score}

The Dice score is a widely used metric in the machine learning and computer vision communities to assess the overlap between two sets of data, making it particularly suitable for evaluating image segmentation methods. In this work, the Dice score was employed to compare the performance of different methods on the various training set compositions depicted in Figure \ref{train}. The Dice score ranges from 0 to 1, where a score of 1 indicates a perfect overlap between the predicted segmentation and the ground truth. In this study, the focus was on accurately classifying pixels as grain boundaries, and thus the Dice score was calculated based on the ability of the models to accurately identify grain boundaries.

\begin{figure}[ht]%
\centering
\includegraphics[width=0.5\textwidth]{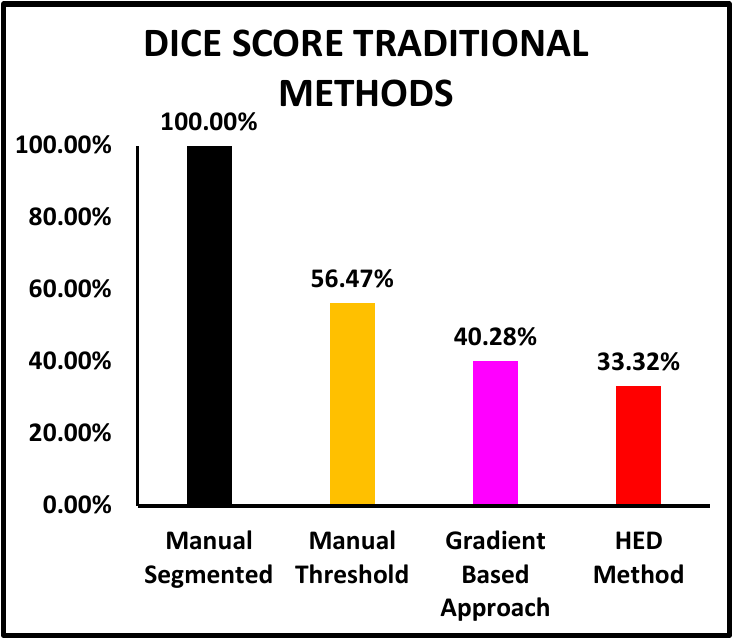}
\caption{Bar graph for the Dice scores for the 3 traditional methods of grain segmentation: Manual Thresholding, Gradient Based Thresholding, and HED method. The ground truth for this measurement is the Manually Segmented images.}
\label{trad}
\end{figure}

\begin{figure}[ht]%
\centering
\includegraphics[width=1\textwidth]{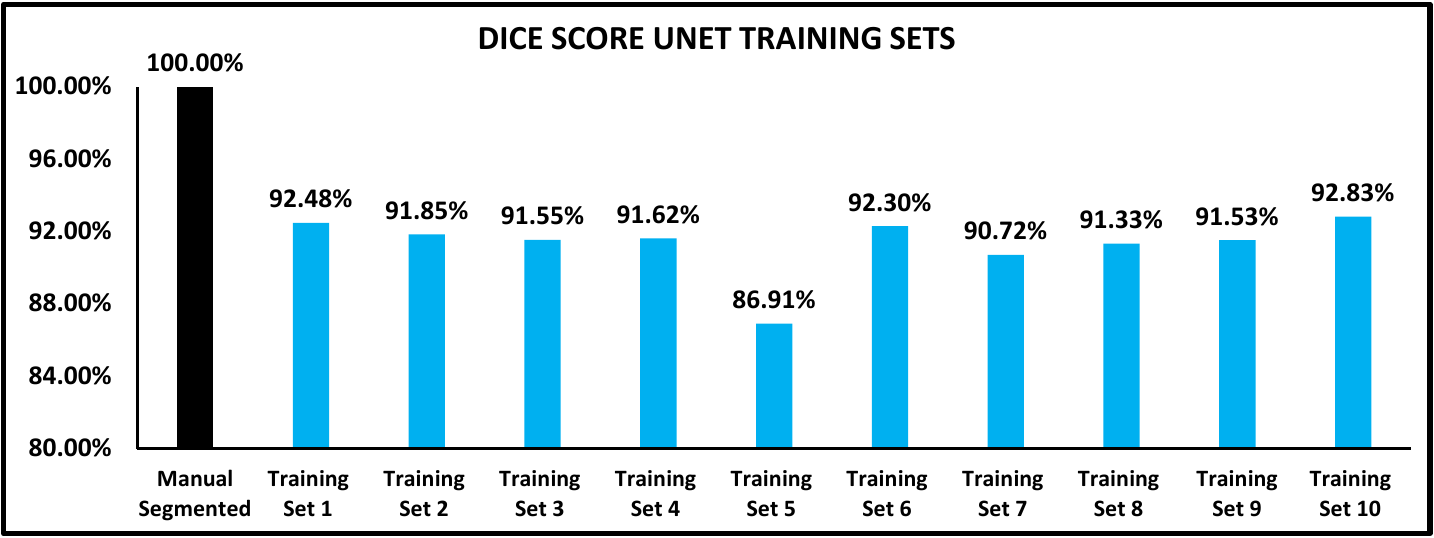}
\caption{Bar graph for the Dice scores for the 10 training sets of data used to train the UNET in this work. The ground truth for this measurement is the Manually Segmented images.}
\label{DICE_UNET}
\end{figure}

The Dice score results are presented in two graphs in this section. Figure \ref{trad} displays the Dice scores for all the traditional methods described in Section 2. Figure \ref{DICE_UNET} depicts the Dice scores for the different U-NET training compositions utilized in this work. It is important to note that the manually segmented image is considered the ground truth in this study, which is why it achieves a perfect score in both figures. The Dice score graphs provide valuable insights into the performance of the segmentation methods and their ability to accurately identify grain boundaries.

In the graph in Figure \ref{trad}, the traditional segmentation methods displayed relatively low Dice scores, with the HED method performing unexpectedly poorly. However, these methods can still be valuable for generating noise to enhance the robustness of machine learning models, as demonstrated in this study. The lower scores emphasize the challenges and limitations of manual and rule-based approaches in accurately capturing complex grain boundary features. The integration of machine learning techniques, like U-NET, offers improved segmentation by capturing both local and global features more effectively.

Figure \ref{DICE_UNET} illustrates that the dice scores for the U-NET models are consistently around 90$\%$. Surprisingly, the inclusion of fully artificially generated data in training sets 2, 3, and 4 did not adversely affect the scores. Even when using solely artificially generated data in training set 5, the dice score only dropped to 87$\%$, which is still remarkable considering the absence of real grain images in the neural network's training.

Furthermore, the dice scores for training sets 6 through 10 demonstrate that incorporating noise generated from the traditional methods does not negatively impact the scores. This suggests that such noise can be utilized to train more robust models that are resistant to the effects of traditional segmentation methods. Notably, training set 10, which included noise from all three traditional segmentation methods, achieved the highest dice score.

\subsection{Grain Data}

\begin{figure}[ht]%
\centering
\includegraphics[width=1\textwidth]{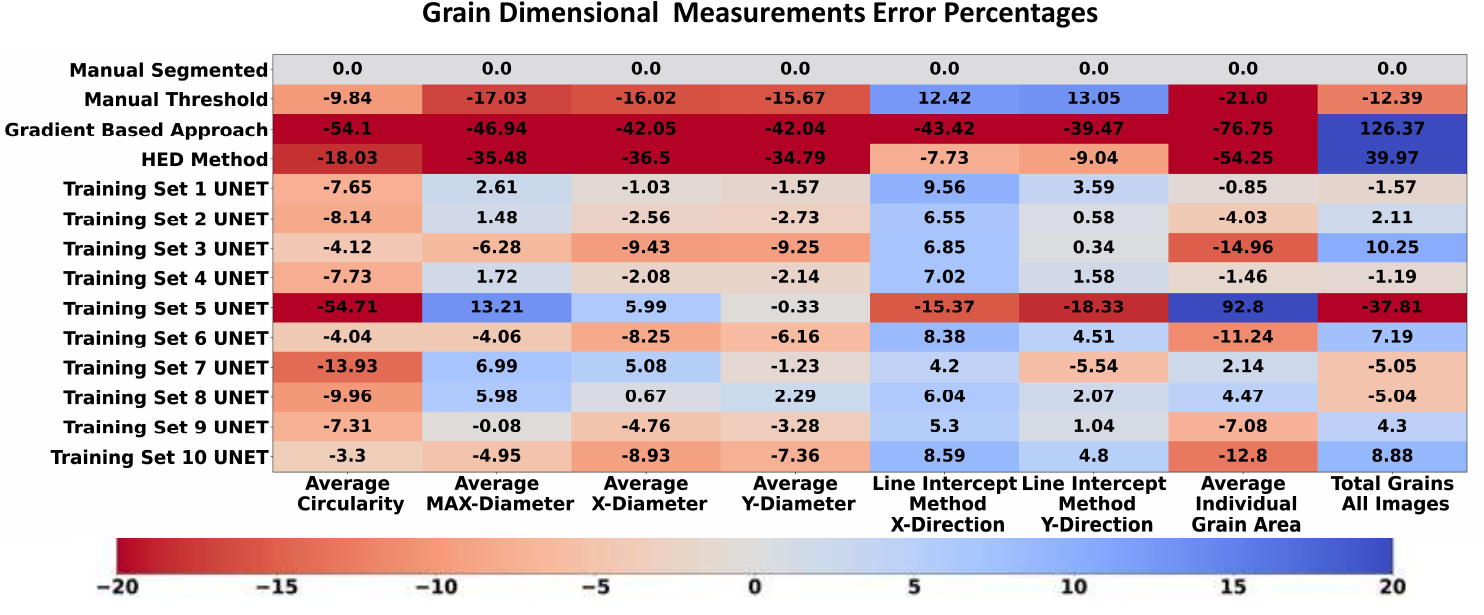}
\caption{A color map superimposed on the numerical values for the various grain dimensional measurement errors. The measurement error is provided for the 3 traditional segmentation methods and all ten of the UNETs trained on differing training sets. The training set compositions are given in Figure \ref{train}. The manually segmented images are regarded as the true value for this error measuremnt.}
\label{grain_error}
\end{figure}

The accuracy of grain geometry measurements is of greater interest to the materials science community than the accuracy of pixel classification. To evaluate the accuracy of the methods used in this paper, the standard error equation (Equation \ref{error}) is employed. The error values are depicted in Figure \ref{grain_error}, where each axis represents a specific grain measurement, such as circularity, average maximum diameter, average X-diameter, average Y-diameter, grain intercept method in the X-direction, grain intercept method in the Y-direction, average grain area, and total grains.

The error values are calculated by comparing the average values obtained from the grain measurements of the different methods to the corresponding average values from the manually segmented images, which are regarded as the ground truth. The colorbar in Figure \ref{grain_error} indicates whether the estimated values overestimate (blue) or underestimate (red) the true values. This analysis provides insights into the accuracy of the grain geometry measurements, highlighting the extent to which the methods over or under predict the values.

\begin{equation}
Error\ Percent = \frac{Measured\ Value - True\ Value}{Measured\ Value}
\label{error}
\end{equation}

Based on the values in Figure \ref{grain_error}, the traditional methods (manual threshold, Gradient Approach, and HED Method) exhibit higher error percentages for all grain measurements compared to the other methods. Among the traditional methods, the manual threshold method shows relatively lower errors. However, both the Gradient Approach and HED Method significantly overestimate the total number of grains in the images, leading to skewed values in other measurements. The average maximum diameter, average X-diameter, and average Y-diameter are also affected by overestimation, although the manual threshold method has lower errors in these measurements compared to the other traditional methods.

In contrast, the UNET training set variations generally exhibit lower errors. Training set 5, trained entirely on artificial data, shows the highest error with an underprediction of grain count by approximately 38$\%$. This is attributed to the presence of incomplete boundaries and small neck regions connecting grains, as evidenced by the lower circularity values. Training set 1 performs well across all measurements, and the inclusion of artificial grains in sets 2, 3, and 4 slightly increases the error percentages. However, these sets outperform the traditional methods and some other training sets. Training set 10, which includes noise from the traditional methods, also demonstrates good performance, indicating the beneficial impact of incorporating diverse noise sources in the training process. Overall, the UNET-based methods outperform the traditional methods, with different training set compositions yielding varying levels of accuracy in grain measurements.

\subsection{Qualitative Results}

The qualitative results provide a closer examination of the approach by presenting individual image results in grain boundary segmentation. This allows for a detailed analysis of the segmentation performance on a case-by-case basis.

\begin{figure}[ht]%
\centering
\includegraphics[width=1\textwidth]{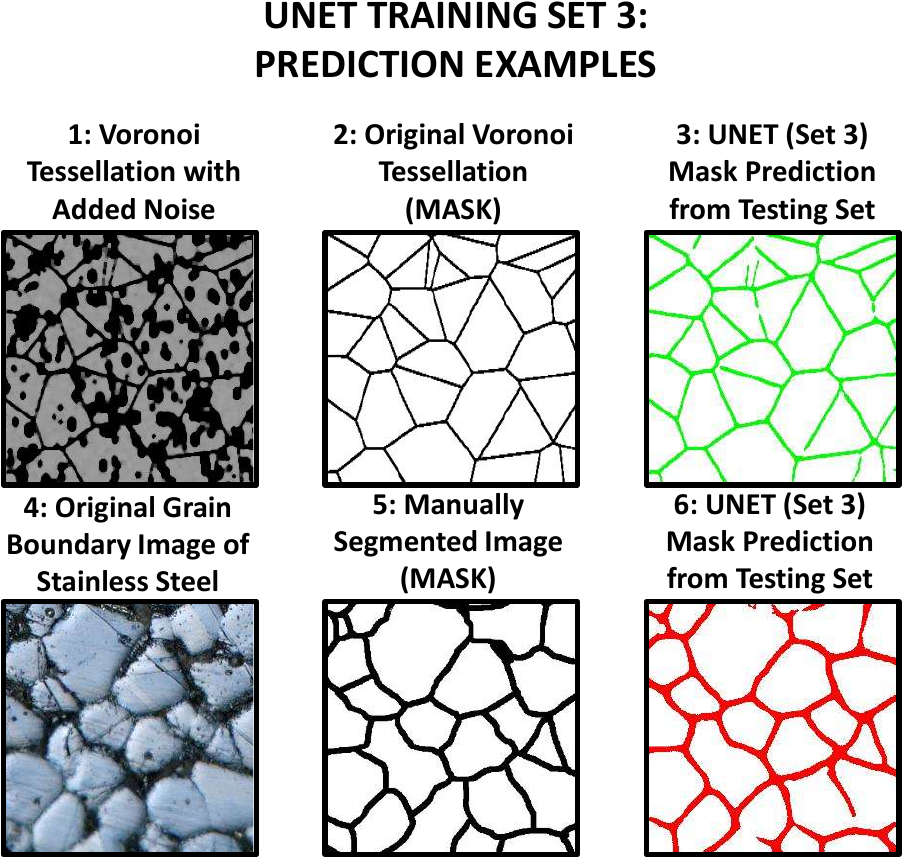}
\caption{Qualitative results from the UNET trained on training set 3. This training set is composed entirely of artificially generated data. (1) An artificial grain generated from the Voronoi Tessellation method. (2) The mask from image 1. (3) Prediction of the mask from UNET trained on data set 3. (4) Real grain image. (5) Mask of the manual segmentation of image 4, complete by a human. (6) UNET training set 3 prediction of the mask of image 4.}
\label{UNET3}
\end{figure}

Figure \ref{UNET3} presents example images from the UNET trained on data set 3, which comprised 50$\%$ real grains with manual segmentations and 50$\%$ artificially generated grains using the Voronoi tessellation method. Image 3 shows the UNET's prediction of the original mask (image 2) based on the input image (image 1) from the testing set. The UNET successfully predicted the mask despite not seeing this specific image during training.

Furthermore, the same UNET trained on data set 3 demonstrated its ability to accurately predict grain boundaries in a real grain image. Image 4 represents a real grain image, image 5 shows the corresponding manually drawn segmentations, and image 6 displays the UNET's prediction of the grain boundaries. The high degree of similarity between images 5 and 6 highlights the exceptional performance of neural networks in grain boundary segmentation tasks. These results showcase the superior capabilities of the UNET model trained on data set 3 in accurately predicting grain boundaries, even when presented with unseen images during training.

\begin{figure}[ht]%
\centering
\includegraphics[width=1\textwidth]{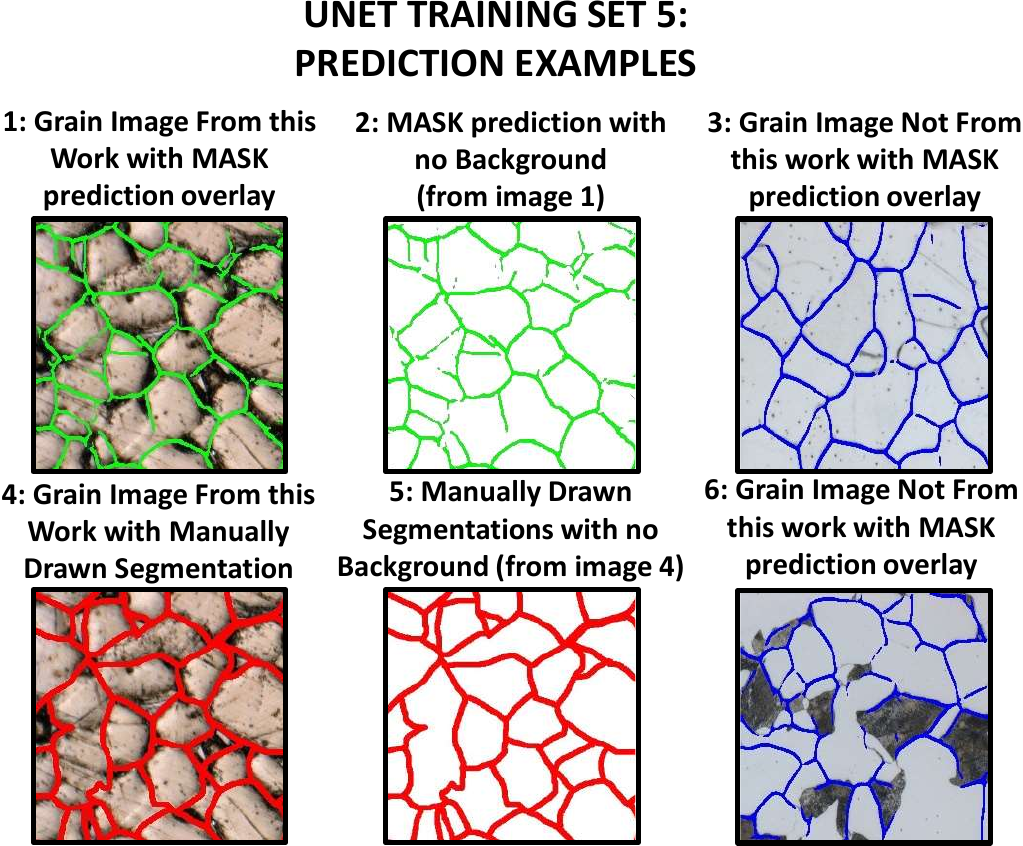}
\caption{Qualitative results from the UNET trained on training set 5. This training set is composed of 50$\%$ manually segmented real grains and 50$\%$ artificially generated grains. (1) Real grain with a prediction of mask from UNET overlaid. (2) The mask from image 1 with background removed. (3) Image of grains this UNET did not train with prediction of mask overlaid (4) Image 1 with manually segmentations overlaid (5) The mask from image 4 with background removed. (6) Image of grains this UNET did not train with prediction of mask overlaid }
\label{UNET5}
\end{figure}

In Figure \ref{UNET5}, the results of a UNET trained exclusively on artificially generated data (training set 5) are presented. This UNET model was never exposed to actual grain boundary images but was solely trained on data generated using the Voronoi tessellation grain image generation method proposed in this work.

Image 1 in Figure \ref{UNET5} displays a real grain image of stainless steel, with the predicted grain boundary mask overlaid in green. Despite not being trained on this specific data, the UNET was able to provide a reasonably accurate estimation of the grain boundaries. For comparison, image 4 shows the manually drawn grain boundaries for the same grain image. Images 2 and 5 show the masks without the grain boundaries, provided for reference. Although the predictions from UNET trained on data set 5 are not perfect, they demonstrate that the model successfully avoided classifying voids and polishing scratches as grain boundaries.

Images 3 and 6 showcase stainless steel images produced through injection molding, which were not used for training and were not captured specifically for this work. These images serve to demonstrate the potential of using artificial data for training more robust models. The UNET trained on data set 5 achieved reasonably accurate grain boundary predictions for these images as well, even in the cases where some grains appeared darker, like in image 6.

\subsection{Summary of Results}

The summary of results from this work will be broken down into the different approaches to keep the statements organized and keep the clarity of each statement intact.

\textbf{The Traditional Methods:} The traditional methods (manual thresholding, gradient-based approach, HED method) require parameter tuning and are not universally applicable. However, they can be used to generate noise for training more robust models.

\textbf{UNET Training Set 1:} Training set 1 performed well on both the dice score and the actual grain measurements, as should be expected. 

\textbf{UNET Training Sets 2-4:} Training Sets 2-4, which included artificially generated grains, maintained high scores in dice score and grain measurement accuracy. Artificial data can effectively increase training data without compromising accuracy.

\textbf{UNET Training Set 5:} This set is composed entirely of artificially generated data, and performed reasonably well, surpassing the traditional methods. The voronoi grain generation method can be used to supplement training data for grain segmentation.

\textbf{UNET Training Set 6-10:} Training Sets 6-10, incorporating noise from traditional methods, achieved high scores in dice score and grain measurement accuracy. Training Set 10, which included noise from all three traditional methods, performed the best. Incorporating noise from these methods enhances model robustness.

\textbf{Planimetric and Line Intercept Method:} Planimetric analysis yielded more accurate grain measurements compared to the line intercept method. Deep learning-based image segmentation can provide suitable images for planimetric analysis.

In summary, the deep learning approach using UNET models demonstrated superior performance compared to traditional methods for grain boundary segmentation. Incorporating artificially generated data and noise from traditional methods can enhance model performance and robustness. Additionally, planimetric analysis provides more accurate grain measurements than the line intercept method.

\section{Conclusion}\label{sec7}

This study has made significant contributions to the field of grain boundary segmentation by introducing a novel method for generating artificial data and demonstrating its potential for improving segmentation accuracy. The generated artificial data incorporated various realistic defects, such as pores, polishing scratches, and small grain fragments, which proved beneficial in training the UNET algorithm to better identify and avoid these imperfections when predicting grain boundaries. This approach addresses a critical challenge faced in grain measurement science, where the availability of annotated real-world data for training segmentation algorithms is often limited.

However, one notable limitation of the UNET trained on the artificial data was the occasional failure to completely enclose the grains in the predicted boundaries, leading to lower dice scores and less accurate grain measurements. Strategies such as refining the generation algorithm, incorporating additional features to guide grain boundary closure, or exploring alternative network architectures may help overcome this limitation and enhance the accuracy of grain boundary segmentation. Despite this limitation, the combination of real and artificial training data showcased promising results, indicating that a diverse training set comprising both real and artificial data can effectively improve grain boundary segmentation. Moreover, the inclusion of noise generated from traditional methods did not compromise segmentation accuracy, suggesting that these noisy data sources can be leveraged to enhance the robustness of the trained models. This finding opens up avenues for future research, where the integration of various data sources, including artificial, real, and noisy data, can lead to the development of more reliable and adaptable machine learning models for grain boundary analysis.

The artificial data generation method presented in this study offers valuable insights and possibilities for advancing grain measurement science. The developed artificial grain generation method can be adapted for different grain structures and offers a quantitative framework for analyzing grain geometry by utilizing the input properties of the Voronoi tessellation pattern. This adaptability allows researchers to explore diverse grain patterns and investigate their influence on material properties. By systematically varying the input properties, valuable insights can be gained into the relationship between grain morphology and material behavior. This approach has the potential to enhance our understanding of structure-property relationships, improve grain boundary segmentation algorithms, and facilitate the design of materials with tailored microstructures.

\section*{Acknowledgements}

We thank the Siemens Energy Center and AM center in North Carolina for Providing the Samples Analyzed in this work

\noindent

\bibliography{sn-bibliography}


\begin{thebibliography}{37}
\ifx \bisbn   \undefined \def \bisbn  #1{ISBN #1}\fi
\ifx \binits  \undefined \def \binits#1{#1}\fi
\ifx \bauthor  \undefined \def \bauthor#1{#1}\fi
\ifx \batitle  \undefined \def \batitle#1{#1}\fi
\ifx \bjtitle  \undefined \def \bjtitle#1{#1}\fi
\ifx \bvolume  \undefined \def \bvolume#1{\textbf{#1}}\fi
\ifx \byear  \undefined \def \byear#1{#1}\fi
\ifx \bissue  \undefined \def \bissue#1{#1}\fi
\ifx \bfpage  \undefined \def \bfpage#1{#1}\fi
\ifx \blpage  \undefined \def \blpage #1{#1}\fi
\ifx \burl  \undefined \def \burl#1{\textsf{#1}}\fi
\ifx \doiurl  \undefined \def \doiurl#1{\url{https://doi.org/#1}}\fi
\ifx \betal  \undefined \def \betal{\textit{et al.}}\fi
\ifx \binstitute  \undefined \def \binstitute#1{#1}\fi
\ifx \binstitutionaled  \undefined \def \binstitutionaled#1{#1}\fi
\ifx \bctitle  \undefined \def \bctitle#1{#1}\fi
\ifx \beditor  \undefined \def \beditor#1{#1}\fi
\ifx \bpublisher  \undefined \def \bpublisher#1{#1}\fi
\ifx \bbtitle  \undefined \def \bbtitle#1{#1}\fi
\ifx \bedition  \undefined \def \bedition#1{#1}\fi
\ifx \bseriesno  \undefined \def \bseriesno#1{#1}\fi
\ifx \blocation  \undefined \def \blocation#1{#1}\fi
\ifx \bsertitle  \undefined \def \bsertitle#1{#1}\fi
\ifx \bsnm \undefined \def \bsnm#1{#1}\fi
\ifx \bsuffix \undefined \def \bsuffix#1{#1}\fi
\ifx \bparticle \undefined \def \bparticle#1{#1}\fi
\ifx \barticle \undefined \def \barticle#1{#1}\fi
\bibcommenthead
\ifx \bconfdate \undefined \def \bconfdate #1{#1}\fi
\ifx \botherref \undefined \def \botherref #1{#1}\fi
\ifx \url \undefined \def \url#1{\textsf{#1}}\fi
\ifx \bchapter \undefined \def \bchapter#1{#1}\fi
\ifx \bbook \undefined \def \bbook#1{#1}\fi
\ifx \bcomment \undefined \def \bcomment#1{#1}\fi
\ifx \oauthor \undefined \def \oauthor#1{#1}\fi
\ifx \citeauthoryear \undefined \def \citeauthoryear#1{#1}\fi
\ifx \endbibitem  \undefined \def \endbibitem {}\fi
\ifx \bconflocation  \undefined \def \bconflocation#1{#1}\fi
\ifx \arxivurl  \undefined \def \arxivurl#1{\textsf{#1}}\fi
\csname PreBibitemsHook\endcsname

\bibitem{do2007effect}
\begin{barticle}
\bauthor{\bsnm{Do~Lee}, \binits{C.}}:
\batitle{Effect of grain size on the tensile properties of magnesium alloy}.
\bjtitle{Materials Science and Engineering: A}
\bvolume{459}(\bissue{1-2}),
\bfpage{355}--\blpage{360}
(\byear{2007})
\end{barticle}
\endbibitem

\bibitem{schempp2013influence}
\begin{barticle}
\bauthor{\bsnm{Schempp}, \binits{P.}},
\bauthor{\bsnm{Cross}, \binits{C.}},
\bauthor{\bsnm{H{\"a}cker}, \binits{R.}},
\bauthor{\bsnm{Pittner}, \binits{A.}},
\bauthor{\bsnm{Rethmeier}, \binits{M.}}:
\batitle{Influence of grain size on mechanical properties of aluminium gta weld
  metal}.
\bjtitle{Welding in the World}
\bvolume{57}(\bissue{3}),
\bfpage{293}--\blpage{304}
(\byear{2013})
\end{barticle}
\endbibitem

\bibitem{wang1995effect}
\begin{barticle}
\bauthor{\bsnm{Wang}, \binits{N.}},
\bauthor{\bsnm{Wang}, \binits{Z.}},
\bauthor{\bsnm{Aust}, \binits{K.}},
\bauthor{\bsnm{Erb}, \binits{U.}}:
\batitle{Effect of grain size on mechanical properties of nanocrystalline
  materials}.
\bjtitle{Acta Metallurgica et Materialia}
\bvolume{43}(\bissue{2}),
\bfpage{519}--\blpage{528}
(\byear{1995})
\end{barticle}
\endbibitem

\bibitem{bai2017effect}
\begin{botherref}
\oauthor{\bsnm{Bai}, \binits{Y.}},
\oauthor{\bsnm{Wagner}, \binits{G.}},
\oauthor{\bsnm{Williams}, \binits{C.B.}}:
Effect of particle size distribution on powder packing and sintering in binder
  jetting additive manufacturing of metals.
Journal of Manufacturing Science and Engineering
\textbf{139}(8)
(2017)
\end{botherref}
\endbibitem

\bibitem{voyiadjis2006transient}
\begin{barticle}
\bauthor{\bsnm{Voyiadjis}, \binits{G.}},
\bauthor{\bsnm{Abed}, \binits{F.}}:
\batitle{Transient localizations in metals using microstructure-based yield
  surfaces}.
\bjtitle{Modelling and Simulation in Materials Science and Engineering}
\bvolume{15}(\bissue{1}),
\bfpage{83}
(\byear{2006})
\end{barticle}
\endbibitem

\bibitem{heo2006influence}
\begin{barticle}
\bauthor{\bsnm{Heo}, \binits{S.}},
\bauthor{\bsnm{Yun}, \binits{J.}},
\bauthor{\bsnm{Oh}, \binits{K.}},
\bauthor{\bsnm{Han}, \binits{K.}}:
\batitle{Influence of particle size and shape on electrical and mechanical
  properties of graphite reinforced conductive polymer composites for the
  bipolar plate of pem fuel cells}.
\bjtitle{Advanced composite materials}
\bvolume{15}(\bissue{1}),
\bfpage{115}--\blpage{126}
(\byear{2006})
\end{barticle}
\endbibitem

\bibitem{uddin2010effect}
\begin{barticle}
\bauthor{\bsnm{Uddin}, \binits{S.M.}},
\bauthor{\bsnm{Mahmud}, \binits{T.}},
\bauthor{\bsnm{Wolf}, \binits{C.}},
\bauthor{\bsnm{Glanz}, \binits{C.}},
\bauthor{\bsnm{Kolaric}, \binits{I.}},
\bauthor{\bsnm{Volkmer}, \binits{C.}},
\bauthor{\bsnm{H{\"o}ller}, \binits{H.}},
\bauthor{\bsnm{Wienecke}, \binits{U.}},
\bauthor{\bsnm{Roth}, \binits{S.}},
\bauthor{\bsnm{Fecht}, \binits{H.-J.}}:
\batitle{Effect of size and shape of metal particles to improve hardness and
  electrical properties of carbon nanotube reinforced copper and copper alloy
  composites}.
\bjtitle{Composites Science and Technology}
\bvolume{70}(\bissue{16}),
\bfpage{2253}--\blpage{2257}
(\byear{2010})
\end{barticle}
\endbibitem

\bibitem{ali2022computational}
\begin{barticle}
\bauthor{\bsnm{Ali}, \binits{H.}},
\bauthor{\bsnm{Stein}, \binits{Z.}},
\bauthor{\bsnm{Fouliard}, \binits{Q.}},
\bauthor{\bsnm{Ebrahimi}, \binits{H.}},
\bauthor{\bsnm{Warren}, \binits{P.}},
\bauthor{\bsnm{Raghavan}, \binits{S.}},
\bauthor{\bsnm{Ghosh}, \binits{R.}}:
\batitle{Computational model of mechano-electrochemical effect of aluminum
  alloys corrosion}.
\bjtitle{Journal of Engineering for Gas Turbines and Power}
\bvolume{144}(\bissue{4}),
\bfpage{041004}
(\byear{2022})
\end{barticle}
\endbibitem

\bibitem{adam20183d}
\begin{barticle}
\bauthor{\bsnm{Adam}, \binits{K.}},
\bauthor{\bsnm{Z{\"o}llner}, \binits{D.}},
\bauthor{\bsnm{Field}, \binits{D.P.}}:
\batitle{3d microstructural evolution of primary recrystallization and grain
  growth in cold rolled single-phase aluminum alloys}.
\bjtitle{Modelling and Simulation in Materials Science and Engineering}
\bvolume{26}(\bissue{3}),
\bfpage{035011}
(\byear{2018})
\end{barticle}
\endbibitem

\bibitem{chen2009modeling}
\begin{barticle}
\bauthor{\bsnm{Chen}, \binits{F.}},
\bauthor{\bsnm{Cui}, \binits{Z.}},
\bauthor{\bsnm{Liu}, \binits{J.}},
\bauthor{\bsnm{Zhang}, \binits{X.}},
\bauthor{\bsnm{Chen}, \binits{W.}}:
\batitle{Modeling and simulation on dynamic recrystallization of 30cr2ni4mov
  rotor steel using the cellular automaton method}.
\bjtitle{Modelling and Simulation in Materials Science and Engineering}
\bvolume{17}(\bissue{7}),
\bfpage{075015}
(\byear{2009})
\end{barticle}
\endbibitem

\bibitem{herriott2019multi}
\begin{barticle}
\bauthor{\bsnm{Herriott}, \binits{C.}},
\bauthor{\bsnm{Li}, \binits{X.}},
\bauthor{\bsnm{Kouraytem}, \binits{N.}},
\bauthor{\bsnm{Tari}, \binits{V.}},
\bauthor{\bsnm{Tan}, \binits{W.}},
\bauthor{\bsnm{Anglin}, \binits{B.}},
\bauthor{\bsnm{Rollett}, \binits{A.D.}},
\bauthor{\bsnm{Spear}, \binits{A.D.}}:
\batitle{A multi-scale, multi-physics modeling framework to predict spatial
  variation of properties in additive-manufactured metals}.
\bjtitle{Modelling and Simulation in Materials Science and Engineering}
\bvolume{27}(\bissue{2}),
\bfpage{025009}
(\byear{2019})
\end{barticle}
\endbibitem

\bibitem{raju2022sintering}
\begin{bchapter}
\bauthor{\bsnm{Raju}, \binits{N.}},
\bauthor{\bsnm{Warren}, \binits{P.}},
\bauthor{\bsnm{Subramanian}, \binits{R.}},
\bauthor{\bsnm{Ghosh}, \binits{R.}},
\bauthor{\bsnm{Fernandez}, \binits{E.}},
\bauthor{\bsnm{Raghavan}, \binits{S.}},
\bauthor{\bsnm{Kapat}, \binits{J.}}:
\bctitle{Sintering behaviour of 3d printed 17-4ph stainless steel}.
In: \bbtitle{Turbo Expo: Power for Land, Sea, and Air},
vol. \bseriesno{86052},
pp. \bfpage{007}--\blpage{17028}
(\byear{2022}).
\bcomment{American Society of Mechanical Engineers}
\end{bchapter}
\endbibitem

\bibitem{van2020roadmap}
\begin{barticle}
\bauthor{\bsnm{Van Der~Giessen}, \binits{E.}},
\bauthor{\bsnm{Schultz}, \binits{P.A.}},
\bauthor{\bsnm{Bertin}, \binits{N.}},
\bauthor{\bsnm{Bulatov}, \binits{V.V.}},
\bauthor{\bsnm{Cai}, \binits{W.}},
\bauthor{\bsnm{Cs{\'a}nyi}, \binits{G.}},
\bauthor{\bsnm{Foiles}, \binits{S.M.}},
\bauthor{\bsnm{Geers}, \binits{M.G.}},
\bauthor{\bsnm{Gonz{\'a}lez}, \binits{C.}},
\bauthor{\bsnm{H{\"u}tter}, \binits{M.}}, \betal:
\batitle{Roadmap on multiscale materials modeling}.
\bjtitle{Modelling and Simulation in Materials Science and Engineering}
\bvolume{28}(\bissue{4}),
\bfpage{043001}
(\byear{2020})
\end{barticle}
\endbibitem

\bibitem{yan2017grain}
\begin{barticle}
\bauthor{\bsnm{Yan}, \binits{F.}},
\bauthor{\bsnm{Xiong}, \binits{W.}},
\bauthor{\bsnm{Faierson}, \binits{E.J.}}:
\batitle{Grain structure control of additively manufactured metallic
  materials}.
\bjtitle{Materials}
\bvolume{10}(\bissue{11}),
\bfpage{1260}
(\byear{2017})
\end{barticle}
\endbibitem

\bibitem{lin2012microstructure}
\begin{bbook}
\bauthor{\bsnm{Lin}, \binits{J.}},
\bauthor{\bsnm{Balint}, \binits{D.}},
\bauthor{\bsnm{Pietrzyk}, \binits{M.}}:
\bbtitle{Microstructure Evolution in Metal Forming Processes}.
\bpublisher{Elsevier},
\blocation{USA}
(\byear{2012})
\end{bbook}
\endbibitem

\bibitem{steinbach2009phase}
\begin{barticle}
\bauthor{\bsnm{Steinbach}, \binits{I.}}:
\batitle{Phase-field models in materials science}.
\bjtitle{Modelling and simulation in materials science and engineering}
\bvolume{17}(\bissue{7}),
\bfpage{073001}
(\byear{2009})
\end{barticle}
\endbibitem

\bibitem{bandyopadhyay2020recent}
\begin{barticle}
\bauthor{\bsnm{Bandyopadhyay}, \binits{A.}},
\bauthor{\bsnm{Zhang}, \binits{Y.}},
\bauthor{\bsnm{Bose}, \binits{S.}}:
\batitle{Recent developments in metal additive manufacturing}.
\bjtitle{Current opinion in chemical engineering}
\bvolume{28},
\bfpage{96}--\blpage{104}
(\byear{2020})
\end{barticle}
\endbibitem

\bibitem{warren2022shrinkage}
\begin{bchapter}
\bauthor{\bsnm{Warren}, \binits{P.}},
\bauthor{\bsnm{Raju}, \binits{N.}},
\bauthor{\bsnm{Krsmanovic}, \binits{M.}},
\bauthor{\bsnm{Ebrahimi}, \binits{H.}},
\bauthor{\bsnm{Kapat}, \binits{J.}},
\bauthor{\bsnm{Subramanian}, \binits{R.}},
\bauthor{\bsnm{Ghosh}, \binits{R.}}:
\bctitle{Shrinkage prediction using machine learning for additively
  manufactured ceramic and metallic components for gas turbine applications}.
In: \bbtitle{Turbo Expo: Power for Land, Sea, and Air},
vol. \bseriesno{85987},
pp. \bfpage{002}--\blpage{05023}
(\byear{2022}).
\bcomment{American Society of Mechanical Engineers}
\end{bchapter}
\endbibitem

\bibitem{raju2021material}
\begin{bchapter}
\bauthor{\bsnm{Raju}, \binits{N.}},
\bauthor{\bsnm{Warren}, \binits{P.}},
\bauthor{\bsnm{Subramanian}, \binits{R.}},
\bauthor{\bsnm{Ghosh}, \binits{R.}},
\bauthor{\bsnm{Raghavan}, \binits{S.}},
\bauthor{\bsnm{Fernandez}, \binits{E.}},
\bauthor{\bsnm{Kapat}, \binits{J.}}:
\bctitle{Material properties of 17-4ph stainless steel fabricated by atomic
  diffusion additive manufacturing (adam)}.
In: \bbtitle{2021 International Solid Freeform Fabrication Symposium}
(\byear{2021}).
\bcomment{University of Texas at Austin}
\end{bchapter}
\endbibitem

\bibitem{vayre2012metallic}
\begin{barticle}
\bauthor{\bsnm{Vayre}, \binits{B.}},
\bauthor{\bsnm{Vignat}, \binits{F.}},
\bauthor{\bsnm{Villeneuve}, \binits{F.}}:
\batitle{Metallic additive manufacturing: state-of-the-art review and
  prospects}.
\bjtitle{Mechanics \& Industry}
\bvolume{13}(\bissue{2}),
\bfpage{89}--\blpage{96}
(\byear{2012})
\end{barticle}
\endbibitem

\bibitem{yakout2018review}
\begin{barticle}
\bauthor{\bsnm{Yakout}, \binits{M.}},
\bauthor{\bsnm{Elbestawi}, \binits{M.}},
\bauthor{\bsnm{Veldhuis}, \binits{S.C.}}:
\batitle{A review of metal additive manufacturing technologies}.
\bjtitle{Solid State Phenomena}
\bvolume{278},
\bfpage{1}--\blpage{14}
(\byear{2018})
\end{barticle}
\endbibitem

\bibitem{frazier2014metal}
\begin{barticle}
\bauthor{\bsnm{Frazier}, \binits{W.E.}}:
\batitle{Metal additive manufacturing: a review}.
\bjtitle{Journal of Materials Engineering and performance}
\bvolume{23},
\bfpage{1917}--\blpage{1928}
(\byear{2014})
\end{barticle}
\endbibitem

\bibitem{tan2020microstructure}
\begin{barticle}
\bauthor{\bsnm{Tan}, \binits{J.H.K.}},
\bauthor{\bsnm{Sing}, \binits{S.L.}},
\bauthor{\bsnm{Yeong}, \binits{W.Y.}}:
\batitle{Microstructure modelling for metallic additive manufacturing: A
  review}.
\bjtitle{Virtual and Physical Prototyping}
\bvolume{15}(\bissue{1}),
\bfpage{87}--\blpage{105}
(\byear{2020})
\end{barticle}
\endbibitem

\bibitem{perera2021optimized}
\begin{barticle}
\bauthor{\bsnm{Perera}, \binits{R.}},
\bauthor{\bsnm{Guzzetti}, \binits{D.}},
\bauthor{\bsnm{Agrawal}, \binits{V.}}:
\batitle{Optimized and autonomous machine learning framework for characterizing
  pores, particles, grains and grain boundaries in microstructural images}.
\bjtitle{Computational Materials Science}
\bvolume{196},
\bfpage{110524}
(\byear{2021})
\end{barticle}
\endbibitem

\bibitem{heyn1925physical}
\begin{bbook}
\bauthor{\bsnm{Heyn}, \binits{E.}}:
\bbtitle{Physical Metallography}.
\bpublisher{Wiley},
\blocation{USA}
(\byear{1925})
\end{bbook}
\endbibitem

\bibitem{subcommittee1996standard}
\begin{bbook}
\bauthor{\bsnm{Subcommittee}, \binits{A.}}:
\bbtitle{Standard Test Methods for Determining Average Grain Size}.
\bpublisher{ASTM International},
\blocation{USA}
(\byear{1996})
\end{bbook}
\endbibitem

\bibitem{PW1}
\begin{botherref}
\oauthor{\bsnm{Warren}, \binits{P.}}:
{Artificial Grains and Real Grains}.
\url{https://www.kaggle.com/datasets/peterwarren/voronoi-artificial-grains-gen}
(2023)
\end{botherref}
\endbibitem

\bibitem{PW3}
\begin{botherref}
\oauthor{\bsnm{Warren}, \binits{P.}}:
{ExONE Stainless Steel 316L Grains 500X}.
\url{https://www.kaggle.com/datasets/peterwarren/exone-stainless-steel-316l-grains-500x}
(2023)
\end{botherref}
\endbibitem

\bibitem{opencv_library}
\begin{botherref}
\oauthor{\bsnm{Bradski}, \binits{G.}}:
{The OpenCV Library}.
Dr. Dobb's Journal of Software Tools
(2000)
\end{botherref}
\endbibitem

\bibitem{PW}
\begin{botherref}
\oauthor{\bsnm{Warren}, \binits{P.}}:
{SinteringTrajectory Github Repository}.
\url{https://github.com/Peterwarren623/GrainBoundaryDetection}
(2022 (accessed Dec 16, 2022))
\end{botherref}
\endbibitem

\bibitem{canny1986computational}
\begin{botherref}
\oauthor{\bsnm{Canny}, \binits{J.}}:
A computational approach to edge detection.
IEEE Transactions on pattern analysis and machine intelligence
(6),
679--698
(1986)
\end{botherref}
\endbibitem

\bibitem{xie2015holistically}
\begin{bchapter}
\bauthor{\bsnm{Xie}, \binits{S.}},
\bauthor{\bsnm{Tu}, \binits{Z.}}:
\bctitle{Holistically-nested edge detection}.
In: \bbtitle{Proceedings of the IEEE International Conference on Computer
  Vision},
pp. \bfpage{1395}--\blpage{1403}
(\byear{2015})
\end{bchapter}
\endbibitem

\bibitem{o2015introduction}
\begin{botherref}
\oauthor{\bsnm{O'Shea}, \binits{K.}},
\oauthor{\bsnm{Nash}, \binits{R.}}:
An introduction to convolutional neural networks.
arXiv preprint arXiv:1511.08458
(2015)
\end{botherref}
\endbibitem

\bibitem{albawi2017understanding}
\begin{bchapter}
\bauthor{\bsnm{Albawi}, \binits{S.}},
\bauthor{\bsnm{Mohammed}, \binits{T.A.}},
\bauthor{\bsnm{Al-Zawi}, \binits{S.}}:
\bctitle{Understanding of a convolutional neural network}.
In: \bbtitle{2017 International Conference on Engineering and Technology
  (ICET)},
pp. \bfpage{1}--\blpage{6}
(\byear{2017}).
\bcomment{Ieee}
\end{bchapter}
\endbibitem

\bibitem{warren2021rapid}
\begin{bchapter}
\bauthor{\bsnm{Warren}, \binits{P.}},
\bauthor{\bsnm{Ali}, \binits{H.}},
\bauthor{\bsnm{Ebrahimi}, \binits{H.}},
\bauthor{\bsnm{Ghosh}, \binits{R.}}:
\bctitle{Rapid defect detection and classification in images using
  convolutional neural networks}.
In: \bbtitle{Turbo Expo: Power for Land, Sea, and Air},
vol. \bseriesno{84966},
pp. \bfpage{004}--\blpage{05013}
(\byear{2021}).
\bcomment{American Society of Mechanical Engineers}
\end{bchapter}
\endbibitem

\bibitem{voronoi1908nouvelles}
\begin{barticle}
\bauthor{\bsnm{Voronoi}, \binits{G.}}:
\batitle{Nouvelles applications des param{\`e}tres continus {\`a} la
  th{\'e}orie des formes quadratiques. deuxi{\`e}me m{\'e}moire. recherches sur
  les parall{\'e}llo{\`e}dres primitifs.}
\bjtitle{Journal f{\"u}r die reine und angewandte Mathematik (Crelles Journal)}
\bvolume{1908}(\bissue{134}),
\bfpage{198}--\blpage{287}
(\byear{1908})
\end{barticle}
\endbibitem

\bibitem{ronneberger2015u}
\begin{bchapter}
\bauthor{\bsnm{Ronneberger}, \binits{O.}},
\bauthor{\bsnm{Fischer}, \binits{P.}},
\bauthor{\bsnm{Brox}, \binits{T.}}:
\bctitle{U-net: Convolutional networks for biomedical image segmentation}.
In: \bbtitle{International Conference on Medical Image Computing and
  Computer-assisted Intervention},
pp. \bfpage{234}--\blpage{241}
(\byear{2015}).
\bcomment{Springer}
\end{bchapter}
\endbibitem

\end{thebibliography}


\end{document}